\documentstyle[twoside,psfig]{article}

\catcode`\@=11
\long\def\@makefntext#1{
\protect\noindent \hbox to 3.2pt {\hskip-.9pt  
$^{{\eightrm\@thefnmark}}$\hfil}#1\hfill}		

\def\@makefnmark{\hbox to 0pt{$^{\@thefnmark}$\hss}}	
	
\def\ps@myheadings{\let\@mkboth\@gobbletwo
\def\@oddhead{\hbox{}
\rightmark\hfil\eightrm\thepage}   
\def\@oddfoot{}\def\@evenhead{\eightrm\thepage\hfil
\leftmark\hbox{}}\def\@evenfoot{}
\def\sectionmark##1{}\def\subsectionmark##1{}}

\oddsidemargin=\evensidemargin
\newcounter{sectionc}\newcounter{subsectionc}\newcounter{subsubsectionc}
\renewcommand{\section}[1] {\vspace{12pt}\addtocounter{sectionc}{1} 
\setcounter{subsectionc}{0}\setcounter{subsubsectionc}{0}\noindent 
	{\tenbf\thesectionc. #1}\par\vspace{5pt}}
\renewcommand{\subsection}[1] {\vspace{12pt}\addtocounter{subsectionc}{1} 
	\setcounter{subsubsectionc}{0}\noindent 
	{\bf\thesectionc.\thesubsectionc. {\kern1pt \bfit #1}}\par\vspace{5pt}}
\renewcommand{\subsubsection}[1] {\vspace{12pt}\addtocounter{subsubsectionc}{1}
	\noindent{\tenrm\thesectionc.\thesubsectionc.\thesubsubsectionc.
	{\kern1pt \tenit #1}}\par\vspace{5pt}}
\newcommand{\nonumsection}[1] {\vspace{12pt}\noindent{\tenbf #1}
	\par\vspace{5pt}}

\newcounter{appendixc}
\newcounter{subappendixc}[appendixc]
\newcounter{subsubappendixc}[subappendixc]
\renewcommand{\thesubappendixc}{\Alph{appendixc}.\arabic{subappendixc}}
\renewcommand{\thesubsubappendixc}
	{\Alph{appendixc}.\arabic{subappendixc}.\arabic{subsubappendixc}}

\renewcommand{\appendix}[1] {\vspace{12pt}
        \refstepcounter{appendixc}
        \setcounter{figure}{0}
        \setcounter{table}{0}
        \setcounter{lemma}{0}
        \setcounter{theorem}{0}
        \setcounter{corollary}{0}
        \setcounter{definition}{0}
        \setcounter{equation}{0}
        \renewcommand{\thefigure}{\Alph{appendixc}.\arabic{figure}}
        \renewcommand{\thetable}{\Alph{appendixc}.\arabic{table}}
        \renewcommand{\theappendixc}{\Alph{appendixc}}
        \renewcommand{\thelemma}{\Alph{appendixc}.\arabic{lemma}}
        \renewcommand{\thetheorem}{\Alph{appendixc}.\arabic{theorem}}
        \renewcommand{\thedefinition}{\Alph{appendixc}.\arabic{definition}}
        \renewcommand{\thecorollary}{\Alph{appendixc}.\arabic{corollary}}
        \renewcommand{\theequation}{\Alph{appendixc}.\arabic{equation}}
        \noindent{\tenbf Appendix \theappendixc #1}\par\vspace{5pt}}
\newcommand{\subappendix}[1] {\vspace{12pt}
        \refstepcounter{subappendixc}
        \noindent{\bf Appendix \thesubappendixc. {\kern1pt \bfit #1}}
	\par\vspace{5pt}}
\newcommand{\subsubappendix}[1] {\vspace{12pt}
        \refstepcounter{subsubappendixc}
        \noindent{\rm Appendix \thesubsubappendixc. {\kern1pt \tenit #1}}
	\par\vspace{5pt}}

\newcommand{\textlineskip}{\baselineskip=13pt}
\newcommand{\smalllineskip}{\baselineskip=10pt}

\def\eightcirc{
\begin{picture}(0,0)
\put(4.4,1.8){\circle{6.5}}
\end{picture}}
\def\eightcopyright{\eightcirc\kern2.7pt\hbox{\eightrm c}} 

\newcommand{\copyrightheading}[1]
	{\vspace*{-2.5cm}\smalllineskip{\flushleft
	{\footnotesize Modern Physics Letters A, #1}\\
	{\footnotesize $\eightcopyright$\, World Scientific Publishing
	 Company}\\
	 }}

\newcommand{\publisher}[2]{{\begin{center}\footnotesize\smalllineskip 
	Received #1\\
	Revised #2
	\end{center}
	}}

\def\abstracts#1#2#3{{
	\centering{\begin{minipage}{4.5in}\footnotesize\baselineskip=10pt
	\parindent=0pt #1\par 
	\parindent=15pt #2\par
	\parindent=15pt #3
	\end{minipage}}\par}} 

\renewenvironment{thebibliography}[1]
	{\frenchspacing
	 \ninerm\baselineskip=11pt
	 \begin{list}{\arabic{enumi}.}
        {\usecounter{enumi}\setlength{\parsep}{0pt}     
	 \setlength{\leftmargin 12.7pt}{\rightmargin 0pt} 
         \setlength{\itemsep}{0pt} \settowidth
	{\labelwidth}{#1.}\sloppy}}{\end{list}}

\newcounter{itemlistc}
\newcounter{romanlistc}
\newcounter{alphlistc}
\newcounter{arabiclistc}

\newcommand{\fcaption}[1]{
        \refstepcounter{figure}
        \setbox\@tempboxa = \hbox{\footnotesize Fig.~\thefigure. #1}
        \ifdim \wd\@tempboxa > 5in
           {\begin{center}
        \parbox{5in}{\footnotesize\smalllineskip Fig.~\thefigure. #1}
            \end{center}}
        \else
             {\begin{center}
             {\footnotesize Fig.~\thefigure. #1}
              \end{center}}
        \fi}

\newcommand{\tcaption}[1]{
        \refstepcounter{table}
        \setbox\@tempboxa = \hbox{\footnotesize Table~\thetable. #1}
        \ifdim \wd\@tempboxa > 5in
           {\begin{center}
        \parbox{5in}{\footnotesize\smalllineskip Table~\thetable. #1}
            \end{center}}
        \else
             {\begin{center}
             {\footnotesize Table~\thetable. #1}
              \end{center}}
        \fi}

\def\@citex[#1]#2{\if@filesw\immediate\write\@auxout
	{\string\citation{#2}}\fi
\def\@citea{}\@cite{\@for\@citeb:=#2\do
	{\@citea\def\@citea{,}\@ifundefined
	{b@\@citeb}{{\bf ?}\@warning
	{Citation `\@citeb' on page \thepage \space undefined}}
	{\csname b@\@citeb\endcsname}}}{#1}}

\newif\if@cghi
\def\cite{\@cghitrue\@ifnextchar [{\@tempswatrue
	\@citex}{\@tempswafalse\@citex[]}}
\def\citelow{\@cghifalse\@ifnextchar [{\@tempswatrue
	\@citex}{\@tempswafalse\@citex[]}}
\def\@cite#1#2{{$\null^{#1}$\if@tempswa\typeout
	{IJCGA warning: optional citation argument 
	ignored: `#2'} \fi}}

\def\pmb#1{\setbox0=\hbox{#1}
	\kern-.025em\copy0\kern-\wd0
	\kern.05em\copy0\kern-\wd0
	\kern-.025em\raise.0433em\box0}

\def\fnt#1#2{\footnotetext{\kern-.3em
	{$^{\mbox{\scriptsize #1}}$}{#2}}}

\def\fpage#1{\begingroup
\voffset=.3in
\thispagestyle{empty}\begin{table}[b]\centerline{\footnotesize #1}
	\end{table}\endgroup}

\def\runninghead#1#2{\pagestyle{myheadings}
\markboth{{\protect\footnotesize\it{\quad #1}}\hfill}
{\hfill{\protect\footnotesize\it{#2\quad}}}}
\font\tenrm=cmr10
\font\tenit=cmti10 
\font\tenbf=cmbx10
\font\bfit=cmbxti10 at 10pt
\font\ninerm=cmr9

\font\eightrm=cmr8

\def\qed{\hbox{${\vcenter{\vbox{			
   \hrule height 0.4pt\hbox{\vrule width 0.4pt height 6pt
   \kern5pt\vrule width 0.4pt}\hrule height 0.4pt}}}$}}

\def\ol{\overline \Lambda}
\def\ms{\overline {\rm MS}}
\def\lms{\Lambda_{\ms}}

\def\ra{\rightarrow}

\def\ran{\rangle}
\def\lan{\langle}
\def\be{\begin{equation}}
\def\ee{\end{equation}}
\def\ba{\begin{eqnarray}}
\def\ea{\end{eqnarray}}
\def\nn{\nonumber}

\begin{document}
\setlength{\textheight}{7.7truein}  

\runninghead{
Mass Effects on the Nucleon Sea Structure Functions
Manuscripts $\ldots$}{
Mass Effects on the Nucleon Sea Structure Functions
Manuscripts $\ldots$}

\normalsize\textlineskip
\thispagestyle{empty}
\setcounter{page}{1}

\copyrightheading{}			

\vspace*{0.88truein}

\fpage{1}
\centerline{\bf
MASS EFFECTS ON THE NUCLEON SEA STRUCTURE FUNCTIONS
}
\vspace*{0.37truein}
\centerline{\footnotesize
Sun Myong Kim
\footnote{
skim@hit.halla.ac.kr
}}
\baselineskip=12pt
\centerline{\footnotesize\it
Department of Liberal Arts and Sciences
}
\baselineskip=10pt
\centerline{\footnotesize\it
Halla University
}
\baselineskip=10pt
\centerline{\footnotesize\it
WonJu, Kangwondo 220-840, Korea
}

\vspace*{0.225truein}

\publisher{(received date)}{(revised date)}

\vspace*{0.21truein}
\abstracts{
Nucleon sea structure functions are studied using
Dokshitzer-Gribov-Lipatov-Altarelli-Parisi (DGLAP) equations with the massive
gluon-quark splitting kernels for strange and charm quarks,
the massless gluon-quark splitting kernels for up and down quarks,
and the massless kernels for all other splitting parts.
The $SU(2)_f$ flavor symmetry for two light quarks,
`up' and `down', is assumed.
Gl\"uck-Reya-Vogt(GRV) and Martin-Roberts-Stirling(MRS) sets are chosen to be
the base structure functions at $Q_0^2=3$
GeV$^2$.
We evolve the sea structure functions
from $Q_0^2=3$ GeV$^2$ to $Q^2=50$ GeV$^2$
using the base structure function sets and DGLAP equations.
Some (about 10\%) enhancement is found in the strange
quark distribution
functions at low $x(<0.1)$ in leading order of the DGLAP equations
compared to results direclty
from those structure function sets at the the value of
$Q^2=50$ GeV$^2$.
We provide the value of $\kappa$ and
also show the behavior of $\kappa(x)=2s(x)/(\bar u(x) + \bar d(x))$
after the evolution of structure functions.
}{}{}



%
%
\vspace*{25pt}
\noindent

The nucleon structure function at low $x$
\cite{bfkl} and at high $x$
may bare non-perturbative effects.
However, we can still investigate the structure function at not so low
$x (\simeq 10^{-1} - 10^{-3})$ with perturbative QCD.
At this low $x$, sea quarks and gluons play important roles in QCD processes.
Disagreements between theoretical and experimental results in various
sum rules suggest the necessity of more improved theoretical analysis
in nucleon structure functions, especially, in sea quark distribution
functions in this region of $x$.

The traditional analysis of sea quark distribution functions
for three light quarks (up, down, and strange)
has been based either on the $SU(3)_f$ flavor symmetry or
on an {\it ad hoc} choice of $s(x)={1\over 2}\bar u(x)={1\over 2}\bar d(x)$
(or $s(x)=\bar u(x)=\bar d(x)$).
Now, we are able to probe these sea structure functions at low
$x$ (up to $\simeq 10^{-4}$) and high $Q^2$ with
better statistics.
The validity of the $SU(3)_f$ symmetry assumption and the {\it ad hoc}
choice of the strange quark structure function has been at stake \cite{cteq}.

Although we believe that there are many signatures that treating
strange quark on an equal footing as other two light quarks is not correct,
it is not clear what physics should be applied and how to implement it to
differentiate its roles in the structure functions.
In this letter, we treat strange and charm quarks massive.
The mass effect is one of several possible factors affecting the structure
functions such as the non-perturbative QCD effect and
the different Pauli exclusion effect for
$\bar u$, $\bar d$, and $s$
quarks which should be taken into account due to
different numbers of valence $u$ and $d$ quarks in the proton.
On this principle, however, there have been speculations that
the effects may be marginal \cite{paulieffect}.

To observe sea quarks in the nucleon (proton), we consider the process
of the deep inelastic scattering of charged lepton off the nucleon.
However, we do not consider the charged weak-current process for simplicity
and clarity \cite{barone}.
To obtain sea structure functions, we use massive kernels \cite{gluck}
for strange and charm quarks in gluon-quark splitting kernels
in Dokshitzer-Gribov-Lipatov-Altarelli-Parisi (DGLAP) equations \cite{dglap}
while we keep the $SU(2)_f$ flavor symmetry for two light quarks, up and down,
by taking massless kernels for the splitting function.
We also keep massless kernels for all other splitting parts.
Unlike our earlier work \cite{us} where we used an approximate expression for 
DGLAP equations for the quark structure functions,
we solve numerically and explicitly
the DGLAP equations for the relevant structure functions
in leading order of $\alpha_s$.

The DGLAP equations and the corresponding kernels to the quark and gluon
distribution functions are given below.
\ba
{dq_i(x,Q^2) \over d\ln Q^2 } &=&
   {\alpha_s (Q^2) \over 2 \pi}
    \int_x^1 {dz\over z}
  \big[ P_{qq}(x/z) ~q_i(z,Q^2)
    +  P_{qg}(x/z) ~g(z,Q^2)
  \big],  \cr
\nn\\
{dg(x,Q^2) \over d\ln Q^2 } &=&
   {\alpha_s (Q^2) \over 2 \pi}
    \int_x^1 {dz\over z} 
  \big[ \sum_i^{N_f} P_{gq}(x/z) ~q_i(z,Q^2)
    + ~P_{gg} (x/z) ~g(z,Q^2)
  \big].
\label{ape}
\ea

Where, the splitting kernels are defined as
\ba
P_{qq}(z)&=& {4\over 3} {1+z^2 \over (1-z)_+} + 2 \delta (1-z), \cr
P_{qg}(z)&=& {1\over 2} \big[ z^2 + (1+z)^2\big], \cr
P^m_{qg}(z)&=& {1\over v} \bigg[ { z^2 + (1+z)^2 \over 2} + {m_q^2\over Q^2}
   {z(3-4z)\over 1-z} -16{m_q^4\over Q^4} z^2\bigg] \cr
  &&-\bigg[ 2{m_q^2\over Q^2}z(1-3z) -8{m_q^4\over Q^4}z^2\bigg]\ln 
  \bigg({1+v\over 1-v}\bigg), \cr
  && {\rm where,~~~} v^2= 1-{4 m_q^2\over Q^2}{z\over 1-z}, \cr
P_{gq}(z)&=& {4\over 3} {1+(1-z)^2 \over z}, \cr
P_{gg}(z)&=& 6 \bigg[ {z \over (1-z)_+} + {1-z \over z} + z(1-z)
               +\big( {11\over 12} - {f\over 18} \big) \delta (1-z)\bigg]. \
\label{kernel}
\ea

Where, $q_i$ is the quark distribution function for the corresponding flavor
`$i$' and `$i$' covers all possible quark and anti-quark flavors.
We include only four quark flavors in this numerical calculation
by excluding bottom and top quarks for simplicity.
The inclusion of these quraks is straight forward but it will not change
our current result noticably.

We use massless kernels, $P_{qq}$, $P_{gg}$, $P_{gq}$ in the
calculation. 
For the gluon-quark splitting kernel, however, we use the massive
kernel $P^m_{qg}$ for strange and charm quarks
while we use the massless kernel $P_{qg}$ for up and down quarks.
Note that $P^m_{qg} \ra P_{qg}$ as $m_q \ra0$ (or $m_q^2/Q^2 \ra0$).
Also, the terms in the massive kernel $P^m_{qg}$
have the power ($m_q^2/Q^2$) suppression
like the ones appearing
in higher twist calculation.
We choose the massive kernel, $P^m_{qg}$, for heavy flavors
only in the $g \rightarrow q$ splitting part
since $g(x,Q^2)$ is the most dominant structure functions at low $x$.
The `+' prescription is defined in standard way,
\ba
\int_0^1 dz {f(z) \over (1-z)_+} \equiv
\int_0^1 dz {f(z)-f(1) \over 1-z}.
\ea

To solve numerically the DGLAP integro-differentioal equations for the quarks
and gluons,
we take the advantage of the differential form of the equations.
We start the evolution of the structure functions at $t=t_0$ and
plug them into the original equations at next $t=t_0+\Delta t$.
Then, iterate them until we get the desired value of $t(Q^2)$.
Where $t=\ln (Q^2/\ol^2)$.
\ba
{dq_i(x,t) \over dt} &=&
   {\alpha_s (t) \over 2 \pi} \int_x^1 {dz\over z}
    \bigg\{ P_{qq} (x/z) \big[ ~q_i (z,t_0) +
            {dq_i(z,t) \over dt}\Big|_{t=t_0} \Delta t ~\big]
 \cr   && \qquad\qquad
~~+ ~P_{qg} (x/z) \big[ ~g(z,t_0) +
              {dg(z,t) \over dt}\Big|_{t=t_0} \Delta t ~\big] \bigg\},
\nn\\
{dg(x,t) \over dt} &=&
   {\alpha_s (t) \over 2 \pi} \int_x^1 {dz\over z}
    \bigg\{ \sum_i^{N_f} P_{gq} (x/z) \big[ ~q_i(z,t_0) +
            {dq_i(z,t) \over dt}\Big|_{t=t_0} \Delta t ~\big]
 \cr   && \qquad\qquad
~~+ ~P_{gg} (x/z) \big[ ~g(z,t_0) +
              {dg(z,t) \over dt}\Big|_{t=t_0} \Delta t ~\big] \bigg\}.
\label{nape}
\ea

As the base structure functions we take two typical structure function
sets (GRV and MRS) \cite{grv}, \cite{mrs} at the initial value, $Q^2_0$
and then, evolve the functions
using the above DGLAP equations to the given value, $Q^2$.
The Gl\"uck-Reya-Vogt(GRV) set contains structure functions
in first two lowest orders in $\alpha_s$.
On the other hand, the Martin-Roberts-Stirling(MRS) set contains the structure
functions
in higher orders in $\alpha_s$ but not in leading order.
The structure functions in the leading order are used in this analysis.
Therefore, for the same order of the calculation, it is more consistent to
use the GRV set for the evolution and compare the evolved result with
the corresponding GRV structure functions.
However, we still include the result using the MRS set for comparision.

We start the evolution at the value of $Q^2_0=3$ GeV$^2$ and end it
at $Q^2=50$ GeV$^2$.
There is no special reason to choose the value of $Q^2_0=3$ GeV$^2$.
This choice of the $Q^2_0$ value is only to get
the non-zero charm distribution function
from the GRV set in the perturbative region.

Our goal is to find the low $x$ behavior of heavy sea quark (strange and charm)
structure funcitons.
We set strange and charm quark masses
to be $m_s=0.2$, $0.5$ GeV and $m_c=1.45$ GeV. 
We also include the evolution result (at $Q^2=50$ GeV$^2$)
with the strange quark mass $m_s=0$ GeV
for the comparision to the strange structure functions of GRV and MRS
at the same $Q^2$.
We choose the value of $\lms$ to be $\lms = 0.2$ GeV for $N_f=4$
which agrees with the value of $\lms$ used in the GRV structure function set
while it is not much different from the one in
the MRS set in which $\lms=0.23$ GeV.

The expression of $v$ in the massive kernel in eq. (\ref{kernel}) contains
the mass threshold (for pair creation process) at parton level.
In other words, $v^2\ge0$ implies that $Q^2 (1-z)/z\ge 4m_q^2$ like that
the Bjorken variable $x=Q^2/[Q^2+(W^2-M^2)]$ should be restricted
by the mass threshold,
$W^2=Q^2 (1-x)/x\ge 4m_Q^2$ for the heavy quark $m_Q$ where
$M$ is the negligible mass of the nucleon and
$W$ is the invariant mass of the virtual photon and the nucleon.

\begin{figure}[h]
\vspace*{13pt}
\centerline{\psfig{file=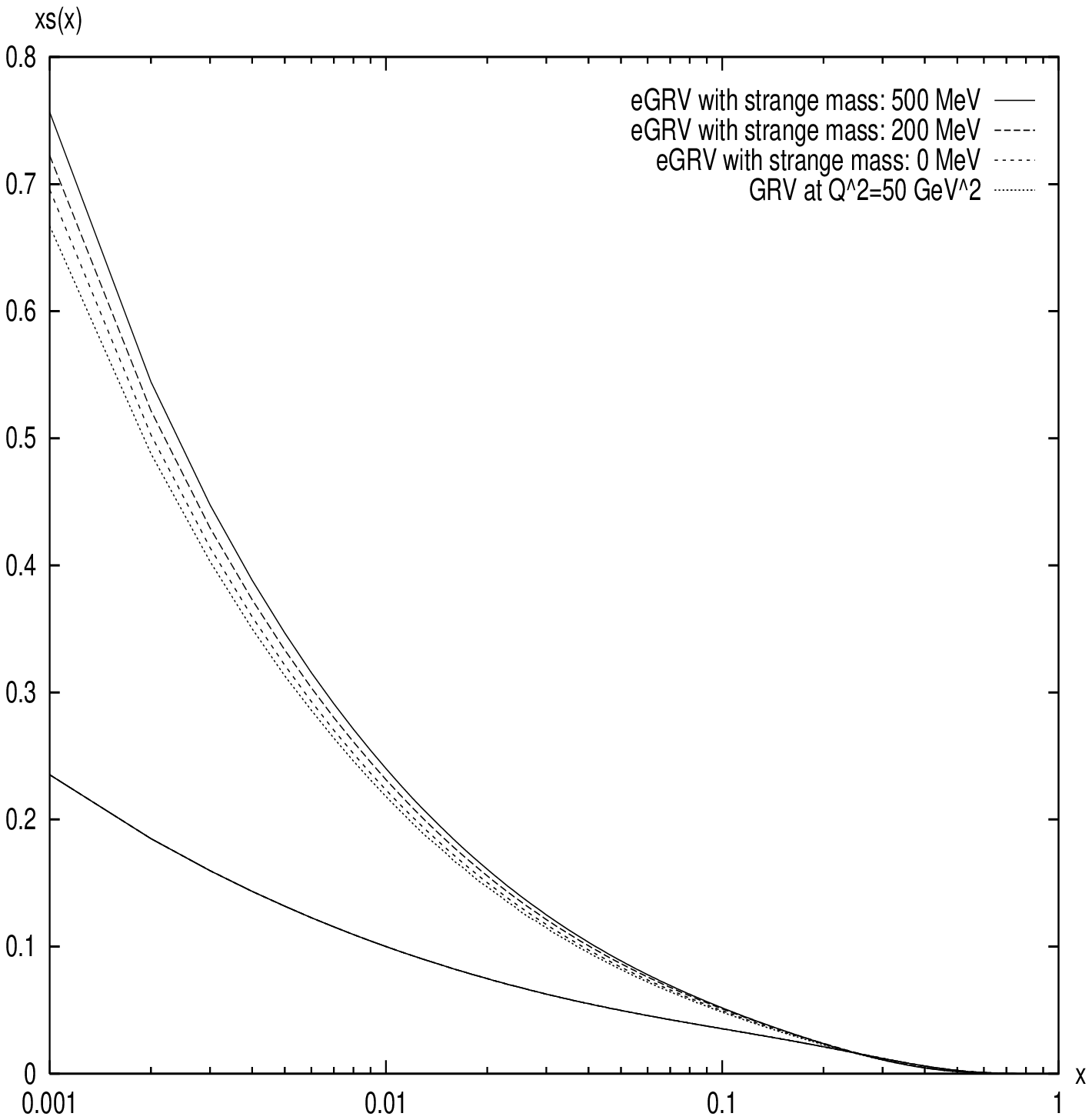}} 
\vspace*{13pt}
\caption{
The strange structure functions are shown here with different strange masses
after the evolution with the GRV base structure function set.
The upper solid, large dashed, small dashed,
dotted, and lower solid curves correspond to the evolved strange structure
functions with strange mass, $m_s=.5$GeV, $m_s=.2$GeV,
$m_s=0$GeV, and the GRV strange structure functions
with $Q^2=50$GeV$^2$ and $Q^2=3$GeV$^2$ respectively.
} 
\label{fig1}
\end{figure}

\begin{figure}[h]
\vspace*{13pt}
\centerline{\psfig{file=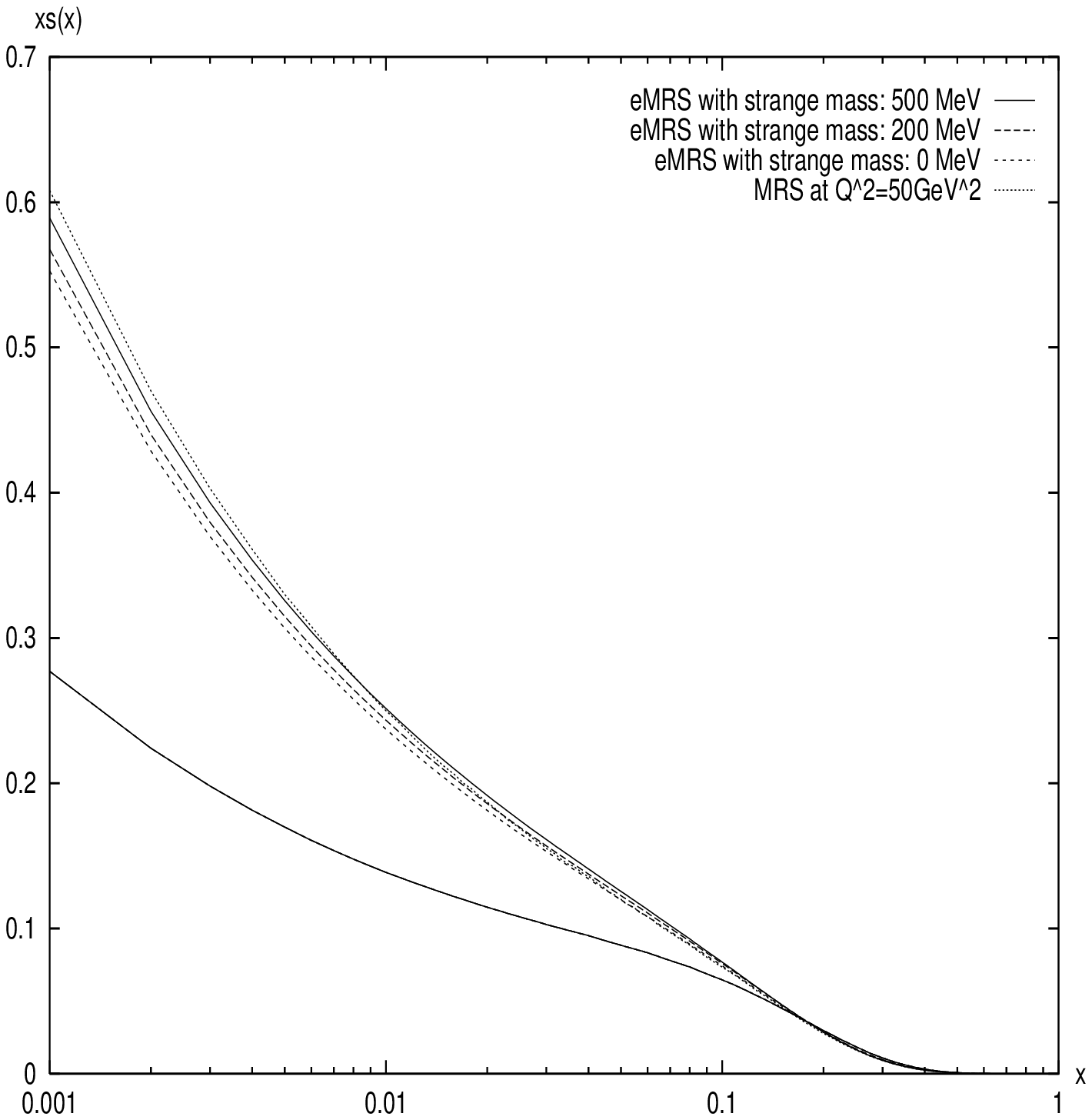}} 
\vspace*{13pt}
\caption{
The strange structure functions are shown here with different strange masses
after the evolution with the MRS base structure function set.
The upper solid, long dashed, short dashed,
dotted, and lower solid curves correspond to the evolved strange structure
functions with strange mass, $m_s=.5$GeV, $m_s=.2$GeV,
$m_s=0$GeV, and the MRS strange structure functions
with $Q^2=50$GeV$^2$ and $Q^2=3$GeV$^2$ respectively.
} 
\label{fig2}
\end{figure}

We observe about 10\% enhancement in the strange quark structure function
at low $x$ and the given $Q^2$.
The boost of this distribution function at such low $x$ is due to the
introduction of the massive kernel in the DGLAP equations.
This massive kernel effect persists for the surprisingly high value of $Q^2$
for the strange quark mass $m_s$.

Due to the complicated dependence of $m_q^2/Q^2$ and $z$,
$P^m_{qg}$ is not so suppressed as was originally thought (See figures 1 and 2).
However, the effect from the massive kernel can not be strong for the
charm quark (See figure 3).
It is fine to use the massive kernel for all range of $Q^2$ that
we considered, $Q^2=3$ GeV$^2 \sim 50$ GeV$^2$.

To show this mass effect more closely,
we also use the massless kernel for the strange quark.
As shown in fig. 1, the result of the evolution with this massless kernel
is smallest and thus closer to the strange quark structure function of
the lowest order GRV structure
function set at $Q^2=50$ GeV$^2$.

Similar behavior can be found in using the MRS structure functions set
(in fig. 2).
The evolved strange structure function with the zero strange mass is smallest
in low $x$
compared to the strange structure functions with the non-zero strange mass.
The difference is that all the obtained strange structure functions
after the evolution are smaller than the one directly from the MRS set.
Remind that the MRS set we used is in next to leading order in $\alpha_s$
while our result is in leading order in $\alpha_s$ like the GRV set used here.
Therefore, the next to leading order correction may be bigger than the mass
effect at low $x$.

We applied the same method to the charm quark structure function (in fig. 3).
We obtained the evolution result which is smaller in using both the GRV set
and the MRS set.
The reason for this behavior is that
the mass threshold of the charm quark pair($c\bar c$), 2.9 GeV,
is big enough to suppress the corresponding structure functions.

\begin{figure}[htbp]
\vspace*{13pt}
\centerline{\psfig{file=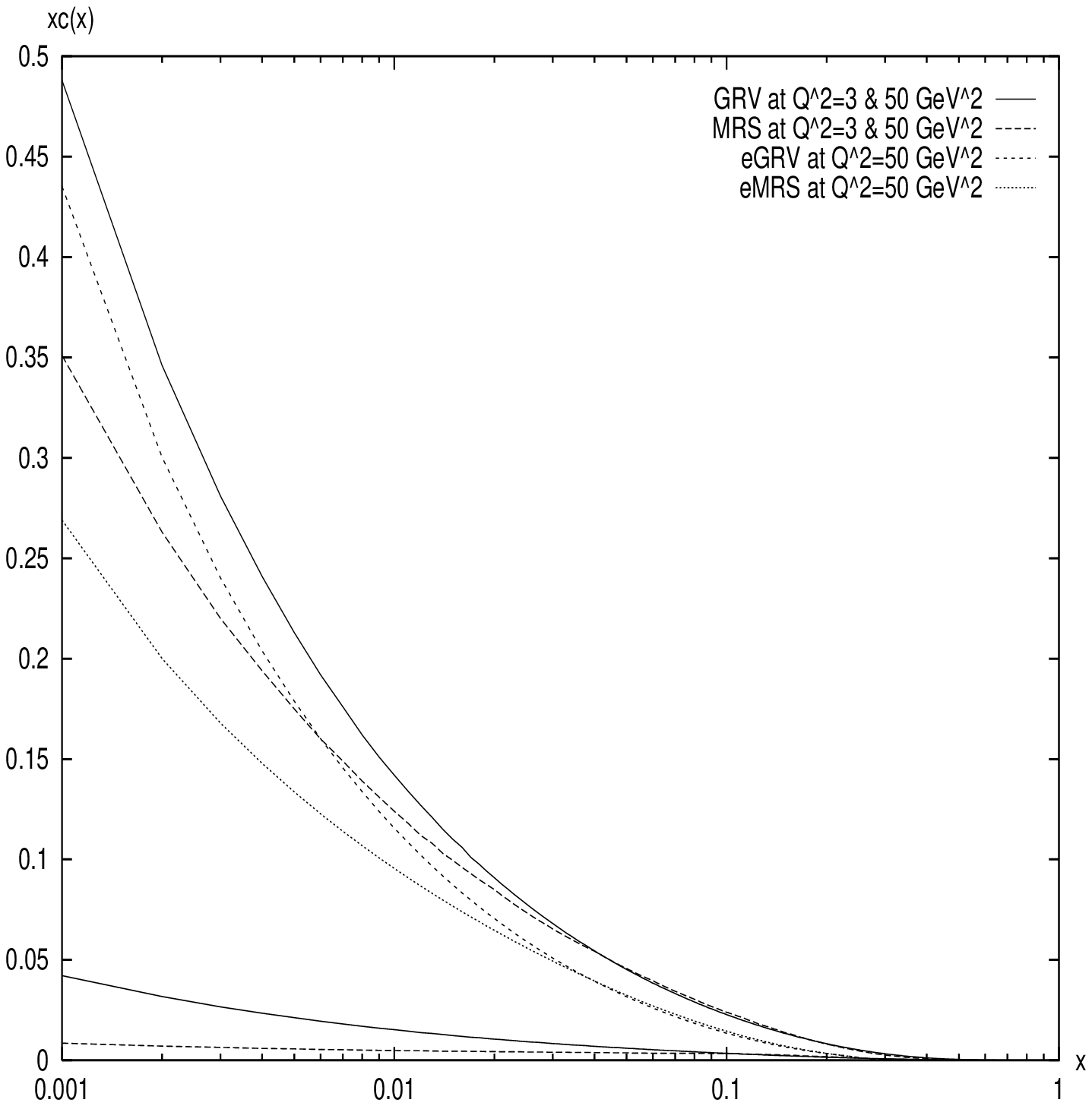}} 
\vspace*{13pt}
\caption{
The charm structure functions are shown here with the mass $m_c=1.45$GeV
for the GRV and the MRS base structure function sets.
The upper solid, upper long dashed, short dashed,
dotted, lower solid, and lower long dashed curves correspond to
the GRV charm structure functions with $Q^2=50$GeV$^2$,
the MRS charm structure functions with $Q^2=50$GeV$^2$,
the evolved charm structure with $Q^2=50$GeV$^2$ and the GRV set,
the evolved charm structure with $Q^2=50$GeV$^2$ and the MRS set,
the GRV charm structure functions with $Q^2=3$GeV$^2$, and
the MRS charm structure functions with $Q^2=3$GeV$^2$
respectively.
} 
\label{fig3}
\end{figure}

The ratio,
$\kappa (x) = 2 s(x) /(\bar u(x) + \bar d(x))$, is also
important in judging the degree of the $SU(3)_f$ symmetry breaking
in terms of variable $x$.
Note the different definition of $\kappa (x)$ from the usual one $\kappa$,
\be
\kappa= {2\lan s(x)\ran\over(\lan\bar u(x)\ran+\lan\bar d(x)\ran)}
={2\int_{x\approx0}^1xs(x)dx\over\int_{x\approx0}^1x\bar u(x)dx
+\int_{x\approx0}^1x\bar d(x)dx}.
\label{kappadef}
\ee
Here, the integration range should be [0,1] instead of [$x$,1].
However, due to the unknown nature of the structure functions at low $x$,
we have to use a low value of $x$ close to `0' but not $x=0$.

$\kappa(x)$ contains some uncertainty at low $x$ like the
structure functions do.
However, this $\kappa(x)$ contains uncertainty also in high value of $x$,
$x>0.5$, in our calculation,
due to very small and thus uncertain values of the sea structure functions
in the range of $x$.
For example, all sea structure functions, $s(x),~\bar u(x)$, and $\bar d(x)$,
are too small with big uncertainty for $x>0.5$ for $\kappa$ to be meaningful.
As shown in the figure 4, $\kappa(x)$ at the high value of $x$ is not reliable.
Fortunately, in calculating $\kappa$, however, those structure functions at
this range of $x$ give the negligible effect.
This uncertainty comes from the technicality in calculation and is 
nothing to do with the nature of QCD unlike the previous uncertainty.

\begin{figure}[htbp] 
\vspace*{13pt}
\centerline{\psfig{file=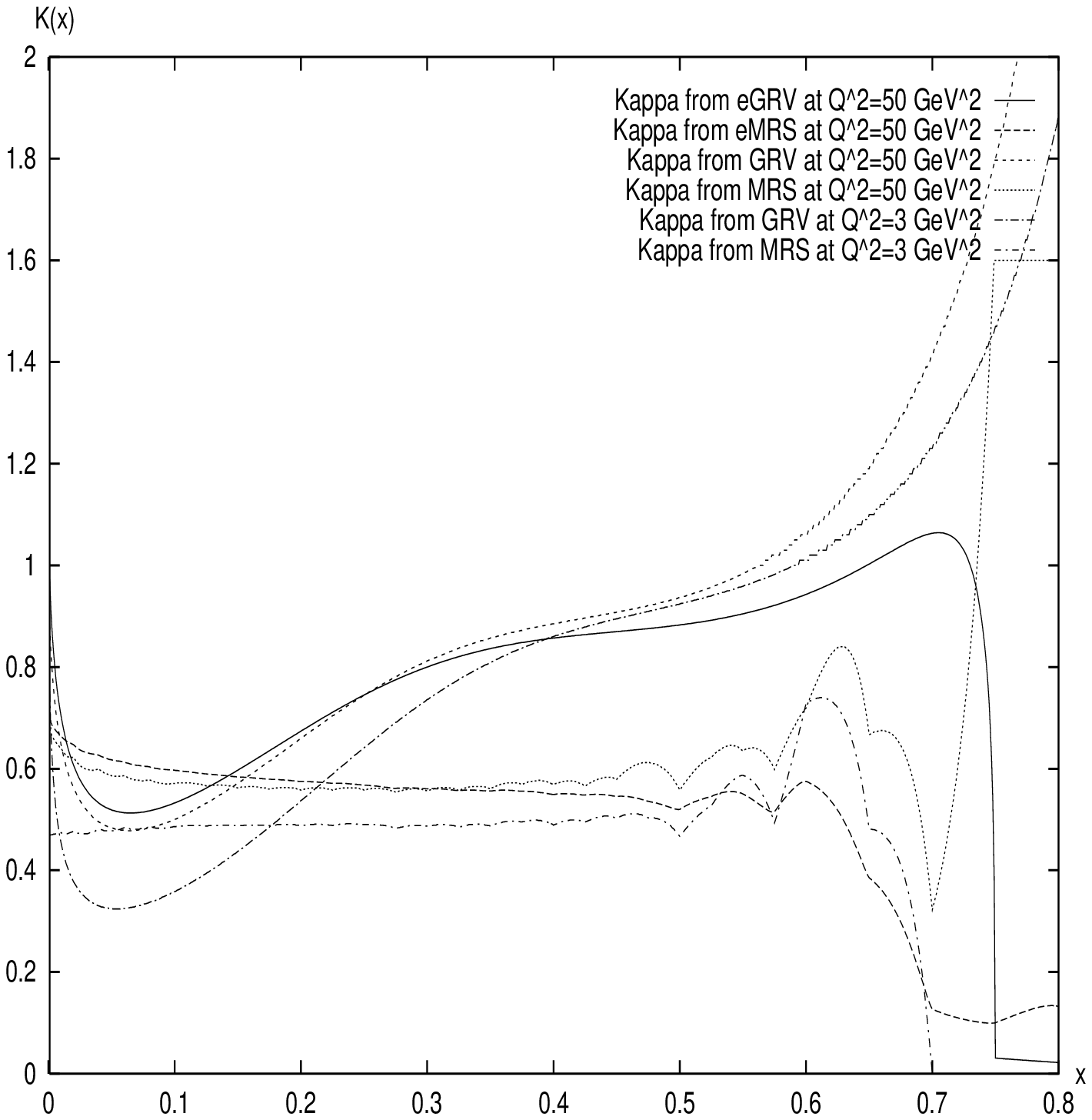}} 
\vspace*{13pt}
\caption{
$\kappa(x)$ in terms of $x$ with two structure function sets and the
evolved structure functions with two different sets.
The solid, long dashed, short dashed,
dotted, long dot-dashed, and short dot-dashed curves correspond to
$\kappa(x)$ from
the evolved structure functions with the GRV set and $Q^2=50$GeV$^2$,
the evolved structure functions with the MRS set and $Q^2=50$GeV$^2$,
the relevant structure functions of the GRV set with $Q^2=50$GeV$^2$,
the relevant structure functions of the MRS set with $Q^2=50$GeV$^2$,
the relevant structure functions of the GRV set with $Q^2=3$GeV$^2$, and
the relevant structure functions of the MRS set with $Q^2=3$GeV$^2$.
}
\label{fig4}
\end{figure}

We found interesting behavior in $\kappa(x)$. $\kappa(x)$ stays almost constant
in $x$ after the use of the MRS set while it changes somewhat (it shoots up
fast at low $x$) after the use of the GRV set as shown in the figure 4.
The value of $\kappa$ also shows the interesting feature.
Values of $\kappa$ both from GRV and MRS sets stay around 0.4 as shown
in the table (\ref{table1}) and increase
slowly for the smaller integration range of $x$
(i.e., bigger value of $x_{max}$ in the table).
The value of $x_{max}$ should be `1'.
Remember the definition of $\kappa$ in eq. (\ref{kappadef}).
We chosed the lower limit of $x$ in the integration to be `0.001'.
On the other hand, $\kappa$'s from our calculation stay around `0.6'.
This is a big difference.
The experimentally measured number of $\kappa$ may bring the way to
improve our analysis further.

\begin{table}[htbp]
\tcaption{
First column, $x_{max}$, stands for the maximum value of the integration.
For example, 0.1 value of $x$ corresponds to
the integration range of $x$ of (0.001, 0.1) and
0.5 to the range of (0.001, 0.5).
Second, third, fourth, and fifth columns in the table stand for
the values of $\kappa$ using
the structure functions directly from the GRV set,
evolved results with the GRV functions,  
the structure functions directly from the MRS set, and
evolved results with the MRS functions, respectively.
In all cases, $Q^2=50$GeV$^2$.
\label{table1}
}

\centerline{\footnotesize\smalllineskip
\begin{tabular}{c c c c c}\\
\hline
$x_{max}$ &GRV &eGRV &MRS &eMRS\\
\hline
0.1 & 0.382 & 0.618 & 0.477 & 0.636  \\
0.2 & 0.390 & 0.614 & 0.479 & 0.629 \\
0.3 & 0.402 & 0.616 & 0.480 & 0.625 \\
0.4 & 0.411 & 0.619 & 0.481 & 0.622 \\
0.5 & 0.419 & 0.621 & 0.481 & 0.621 \\
0.6 & 0.424 & 0.623 & 0.481 & 0.620 \\
0.7 & 0.428 & 0.624 & 0.481 & 0.620 \\
0.8 & 0.430 & 0.625 & 0.482 & 0.619 \\
0.9 & 0.432 & 0.626 & 0.482 & 0.619 \\
\hline\\
\end{tabular}}
\end{table}

\nonumsection{Acknowledgments}
\noindent
This work was supported in part by the School Research Fund
at Halla Institute of Technology. I would like to thank to YVRC
at Yonsei University
for the hospitality.

\nonumsection{References}

\end{document}